# Modulation of Negative Index Metamaterials in the Near-IR Range

Evgenia Kim[(1)†], Wei Wu[(2)], Ekaterina Ponizovskaya[(2)],
Zhaoning Yu[(2)], Alexander M. Bratkovsky[(2)], Shih-Yuang Wang[(2)], R. Stanley Williams[(2)], and Y. Ron Shen[(1)]

[(1)] Department of Physics, University of California, Berkeley, California 94720

[(2)] Quantum Science Research, Hewlett-Packard Laboratories, Palo Alto, California 94304

## Abstract

Optical modulation of the effective refractive properties of a "fishnet" metamaterial with a Ag/Si/Ag heterostructure is demonstrated in the near-IR range and the associated fast dynamics of negative refractive index is studied by pump-probe method. Photo excitation of the amorphous Si layer at visible wavelength and corresponding modification of its optical parameters is found to be responsible for the observed modulation of negative refractive index in near-IR.

[†] Email: evgenia_kim@berkeley.edu

Negative index metamaterials (NIM) that exhibit unique refractive properties [1-3] are currently a focus of research in optoelectronics. Numerous unusual properties and applications of NIM have been discussed [4-6]. Some of them on optical communications and data processing, such as near-field photonic links, require modulation of the effective negative refractive index of the material. This can be achieved by modulating the optical properties of the constituents of an NIM. We report here the first study of optical modulation of an NIM in the near-IR range. Using the pump/probe method, we observed a pump-induced change of 40% in the effective negative refractive index of an NIM composed of a Ag/Si/Ag fishnet heterostructure, with a relaxation time of 58 ps limited by carrier relaxation in Si.

In this study, the fishnet structure [7] was designed with the use of FDTD method (see Fig. 1a) to have a negative refractive index in the 1.6-1.8 μm near-IR range (Fig. 1d) and a magnetic resonance around 1.7 μm (Fig. 1c). The fishnet on a glass substrate consisted of two 25 nm Ag metallic layers separated by a 80 nm amorphous(α−) Si layer and perforated by a periodic array of holes (Fig. 1 b). The period of the resulting network of metallic wires was 320 nm along the wires in both directions (Fig. 1a). The widths of the Ag wires along the two perpendicular directions were ~220 nm and ~110 nm for the bottom layer, respectively, and were approximately 40% smaller for the top layer as a result of our fabrication procedure (Fig. 1b). The fishnet was fabricated by using nanoimprint and electron-beam lithography. The fabrication procedure was described in detail elsewhere [8]. Figure 1e shows the SEM image of the sample. A similar sample, with $SiO_2$ silica replacing Si as the dielectric spacer, was studied earlier to demonstrate the effectiveness of such fishnet structures as negative index materials in the near-IR range [9].

Linear transmission and reflection spectra were measured in order to characterize the sample. The measurements have been carried out with a Nd:YAG laser/optical parametric system generating 20 ps pulses tunable in the entire near-IR range. The experimental setup was the same as that described earlier [8]. For transmission measurement, the input beam was normal to the surface with polarization parallel to the thin Ag wires. For reflection measurement, the same polarized beam was tilted by $10^o$ to the normal to the surface. The phase difference was also measured in transmission and reflection of the two beams polarized parallel to the thin and thick Ag wires, respectively. To study the pump-induced change of transmission and reflection, we adopted the pump/probe method using Q-switched YAG:$Nd^{3+}$ doubled output at wavelength of 532 nm as the pump pulses and the time-delayed wavelength tunable IR pulses from OPO system as the probe. We expected that the pump would excite carriers in Ag and Si, modify their refractive indices, and alter the optical responses of the fishnet.

Displayed in Figs. 2a and 2b are the transmittance and reflectance spectra for our sample, with and without pump. They show a resonant structure at ~1.7 μm in agreement with the theoretical prediction presented in Figs. 1b and 1c. The pump fluence was 320 $\mu J/cm^2$. It is seen that the pump induces a red shift of 15±2 nm, and a decrease of ~50% in the peak magnitude of the magnetic resonance in the transmittance spectrum. Results of the phase measurement are shown in Fig. 2c. The observed phase difference is mainly due to the effect of the magnetic resonance seen only by the waves with polarization of magnetic field vector along the thick Ag wires. Without the pump, it reaches $-38°$ in transmission and $60°$ in reflection at the peak of the resonance. With the pump, the values change to $-25°$ and $42^o$, respectively, following the change of the resonance structure.

Real and imaginary parts of the effective refractive index, *n*, can then be deduced from the experimental data in Figc. 2a and 2c using the method of Ref.[10], with results shown in Fig. 2d. At the magnetic resonance, $\mathrm{Re}(n_{eff})$ for fishnet exhibits a dip reaching a value of $\mathrm{Re}(n_{eff}) = -2.4$. With the pump reducing the resonance strength, it changes to $\mathrm{Re}(n_{eff}) = -1.5$. Fig. 2e also shows the measured transmittance at the magnetic resonance as a function of the pump fluence. The linear relation indicates that the effect is due to a linear increase of pump absorption in the structure.

Pumping the sample (i) produces free carriers in Si and Ag and their relaxation also leads to (ii) heating of the sample. Both processes could modify the optical constants of the materials and hence the optical response of the metamaterial, but carrier relaxation is expected to be much faster than heating. In our pump/probe measurement, we measured a set of transmission spectra at different delay times between pump and probe, and observed the recovery of the induced changes on the resonance structure. Fig. 3a shows the pump-induced change of transmission at the resonance peak of the fishnet as a function of the probe time delay for the pump fluence of 320 μJ/cm$^2$. For comparison, we also display in Fig. 3b the cross-correlation trace of our pump and probe pulses obtained from sum-frequency generation in a barium borate crystal.

To fit the experimental data for the fishnet in Fig. 3a one can use the following expression:

$$S \propto \int_{-\infty}^{\infty}\int_{-\infty}^{\infty} \theta(t_2 - t_1)\cdot[e^{-a(t_2-t_1)} + A]\cdot e^{-\frac{2(t_2-\tau)^2}{w^2}} e^{-\frac{2t_1^2}{w^2}} dt_1 dt_2 . \qquad (1)$$

Here, $\theta(t_2 - t_1)$ is the step function, the exponential decay term exp[-α($t_2$-$t_1$)] and the constant *A* describe, respectively, the effects of carrier relaxation and heating on modulation; the Gaussian functions represent the pump and probe pulse profiles that reproduce the cross-correlation trace in Fig. 3b with *w* = 19 ps, and τ is the time delay between pump and probe pulses. Figure 3a shows a fit of Eq. (1) to experimental data with τ = 58 ps.

We expect that the pump-induced modulation comes mainly from photo-excitation of carriers in the α–Si layer. In that case, the observed fast relaxation would reflect the carrier relaxation in the Si layer. A similar pump/probe measurement are carried out on an 80-nm α–Si film alone, without top and bottom Ag layers The pump-induced absorption and reflectance changes versus time for α–Si film are shown in Fig. 3a and 3b, respectively. The initial dip in reflectance results from a pump-induced reduction of the refractive index, and can be associated with pump-induced free carriers in α–Si. The change in the absorption (in Fig. 3a) is determined from simultaneous measurements of reflectance and transmittance. The decay of absorption changes can be fit by a single exponential with a relaxation time of 50 ps (dashed line in Fig.3 a), characteristic of carrier relaxation in α–Si [11, 12]. The fact that the pump-induced modulation of our fishnet sample has a decay closely resembling that of the free α-Si film indicates that free carrier excitation and relaxation is indeed the dominant mechanism responsible for the modulation, while the excitations in the Ag wires appear not to be important. The tail observed at a long probe delay time, on the other hand, must have resulted from a thermal modulation as the excited carriers relaxed and released the energy to heat up the sample.

To further confirm that carrier excitation in α-Si is the dominating mechanism underlying the observed pump-induced modulation of the fishnet structure, we deduced from the pump/probe measurement of the α-Si film (without silver layers) a maximum pump-induced refractive index change of $\Delta n_{\mathrm{Si}} = \Delta n_{\mathrm{Si}}' + \Delta n_{\mathrm{Si}}'' = -0.055 \pm 0.01 + i(0.02 \pm 0.005)$ for a pump fluence of 320 μJ/cm$^2$. The imaginary part of the index is due to finite conductivity of photoinduced carriers, that we estimated to be about $\sigma_{ph} = 1.3 \times 10^{13}\,\mathrm{s}^{-1}$. The FDTD calculation of the effective refractive index of the fishnet structure with the changes in α–Si refractive index $\Delta n_{\mathrm{Si}} = -0.055 + i0.008)$ and $\Delta n_{\mathrm{Si}} = -0.055 + i0.048$ shows the red shift of the magnetic resonance [the dip in $\mathrm{Re}(n_{eff})$] of about 5 and 30 nm and decreases of the resonant amplitude by 30% and 70%, respectively, Fig. 4. The results are in fair agreement with the experimental observation shown in Fig. 2d. The pump/probe measurement on a fishnet structure with silica replacing Si are made and found that the maximum pump-induced change of transmittance with 320 μJ/cm$^2$ is ≤5%. Because silica does not absorb at the pump frequency, the effect, if any, would have come from pump excitation of the Ag wires. This again indicates that modulation of the fishnet structure by excitation of the Ag wires is not effective.

In conclusion, pump-probe experiments and FDTD simulations demonstrate photo-induced modulation of the effective negative refractive index of a Ag/Si/Ag fishnet structure. A pump with fluence of 320 μJ/cm$^2$ at visible wavelength can change the refractive index of a Ag/Si/Ag fishnet negative index structure at the resonance from $n_{eff} = -2.4 + i1.7$ to $n_{eff} = -1.5 + i1.5$. Photoinduced carriers in the α–Si spacer are responsible for the modulation. It is characterized by dynamic response of 58 ps governed by the carrier relaxation time in α–Si. The present work opens up the possibility of fast switching and/or modulation of NIM devices in the optical range.

The authors thank DARPA for partial support.

**References**

1. L.I. Mandelshtam, *Lectures in Optics, Relativity, and Quantum Mechanics* (Moscow, Nauka, 1972), p. 389.

2. R.A. Silin, Usp. Fiz. Nauk, **175**, 562 (2006); R.A. Silin and V.P. Sazonov, *Delay Systems* (Radio, Moscow, 1966).

3. V.G. Veselago, Usp. Fiz. Nauk, **92**, 517 (1967) [Sov. Phys. Usp. **10**, 509 (1968).]

4. J.B. Pendry, Phys. Rev. Lett. **85,** 3966-3969 (2000).

5. N. Fang, H. Lee, X. Zhang, Science, 308, 534-537 (2005).

6. D.O.S. Melville, R.J. Blaikie, C. R. Wolf, Appl. Phys. Lett. **84**, 4403 (2004).

7. S. Zhang, W. Fan, B. K. Minhas, A. Frauenglass, K. J. Malloy, and S. R. J. Brueck S, Phys. Rev. Lett. **94**, 037402 (2005).

8. W. Wu, Y. Liu**,** E.M. Kim, Z. Yu, N. Fang, X. Zhang, Y.R. Shen, S.Y. Wang, App. Phys. Lett., **90**, 063107 (2007).

**9.** E.M. Kim, W. Wu**,** E**.** Ponizovskaya, Y. Liu, Z. Yu, A.M. Bratkovsky, N. Fang, X. Zhang, Y.R. Shen, and S.Y. Wang, Appl. Phys. A **87**, 143 (2007).

**10.** D. Smith, S. Schultz, P. Markos, and C. Soukoulis, Phys. Rev. B **65**, 195104 (2002).

**11.** A. Esser, K. Seibert, H. Kurz, G.N. Parsons, C. Wang, B.N. Davidson, G. Lucovsky, and R.J. Nemanich, Phys. Rev. B **41,** 2879 (1990).

**12.** M. Kubinyi, A. Grofcsik, and W. Jones, J. Molec. Struc. **408**, 121 (1997).

**Figure captions**

Figure 1. Schematic of the "fishnet" Ag/Si/Ag structure: (a) top view, (b) side view. (c) Effective magnetic permeability $\mu_{eff}$ and (d) effective refractive index $n_{eff}$ for the "fishnet" structure calculated by the FDTD method with the structural parameters given on panel (b), with pump off. (e) SEM images of a fabricated fishnet structure: the top frame has a lower magnification than the bottom frame.

Figure 2. (a) Transmittance and (b) reflectance spectra from the "fishnet" structure without pump (black dots) and with a pump fluence of 320 $\mu$J/cm$^2$ (open black dots) at zero pump-probe time delay. (c) Phase difference spectra for transmitted and reflected light without the pump (black and red dots, respectively) and with the pump (open black and red dots, respectively). (d) Real and imaginary parts of the refractive index deduced from experimental data without the pumping (black and red dots, respectively) and with the pumping (open black and red dots, respectively). (e) Variation of transmittance at the magnetic resonance as a function of the pump fluence.

Figure 3. (a) Pump-induced transmission change at the magnetic resonance of the fishnet (red dots) and pump-induced absorption variation from an amorphous (α−) Si film (black open dots) as functions of the probe time delay for a pump fluence of 320 μJ/cm$^2$ (b) Cross-correlation trace of pump and probe pulses obtained from sum-frequency generation in a barium borate crystal (red dots) and the time-resolved pump-induced reflectance change from an amorphous (α−) Si film (open black dots).

Figure 4.
FDTD simulation of effective refractive indices of the fishnet structure with various refractive index changes in α−Si layer: $\Delta n_{\mathrm{Si}} = 0$ (solid and dashed-dotted curves), $\Delta n_{\mathrm{Si}} = -0.055 + i\ 0.008$ (dashed and shot dashed curves) that corresponds to conductivity of Si $\sigma_{ph}=5\cdot10^{12}$ s$^{-1}$, $\Delta n_{\mathrm{Si}} = -0.055 + i\ 0.048$ (dotted and shot dotted curves) with $\sigma_{ph}=3\cdot10^{13}$ s$^{-1}$, and $\Delta n_{Si} = -0.055 + i \cdot 0.16$ with $\sigma_{ph}=10^{14}$ s$^{-1}$ (shot dotted and shot dashed-dotted)

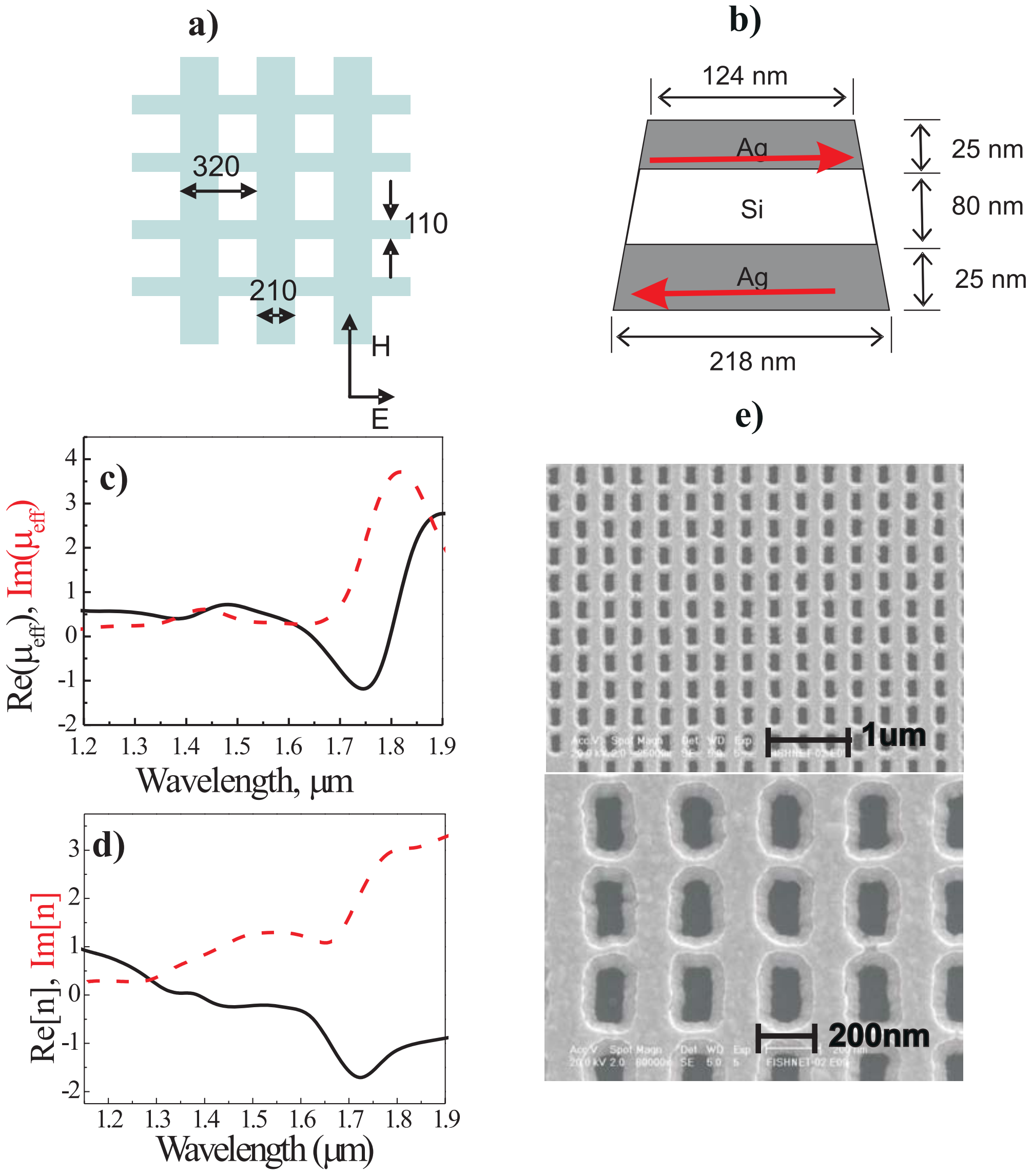

**Figure 1.**

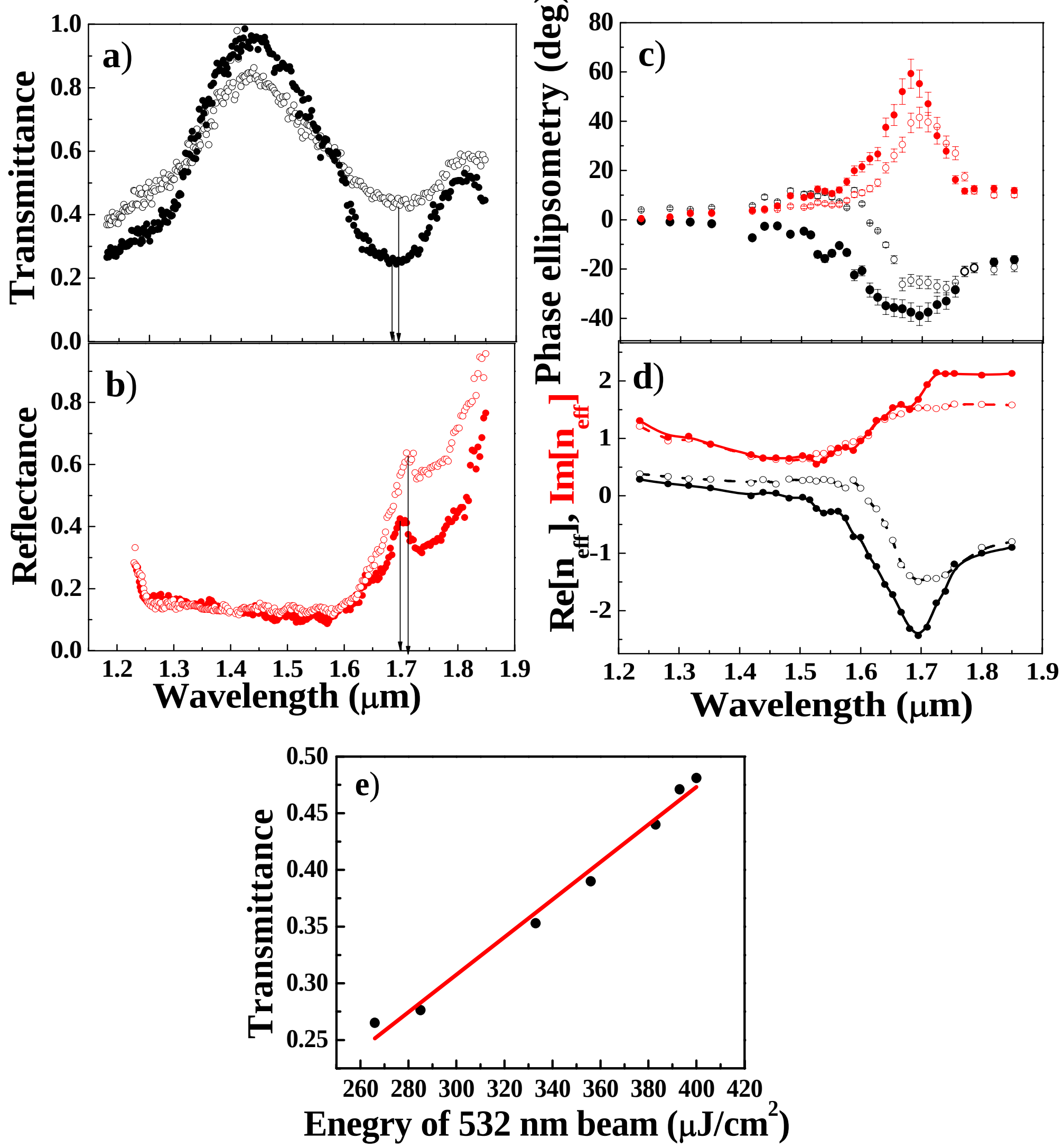

**Figure 2.**

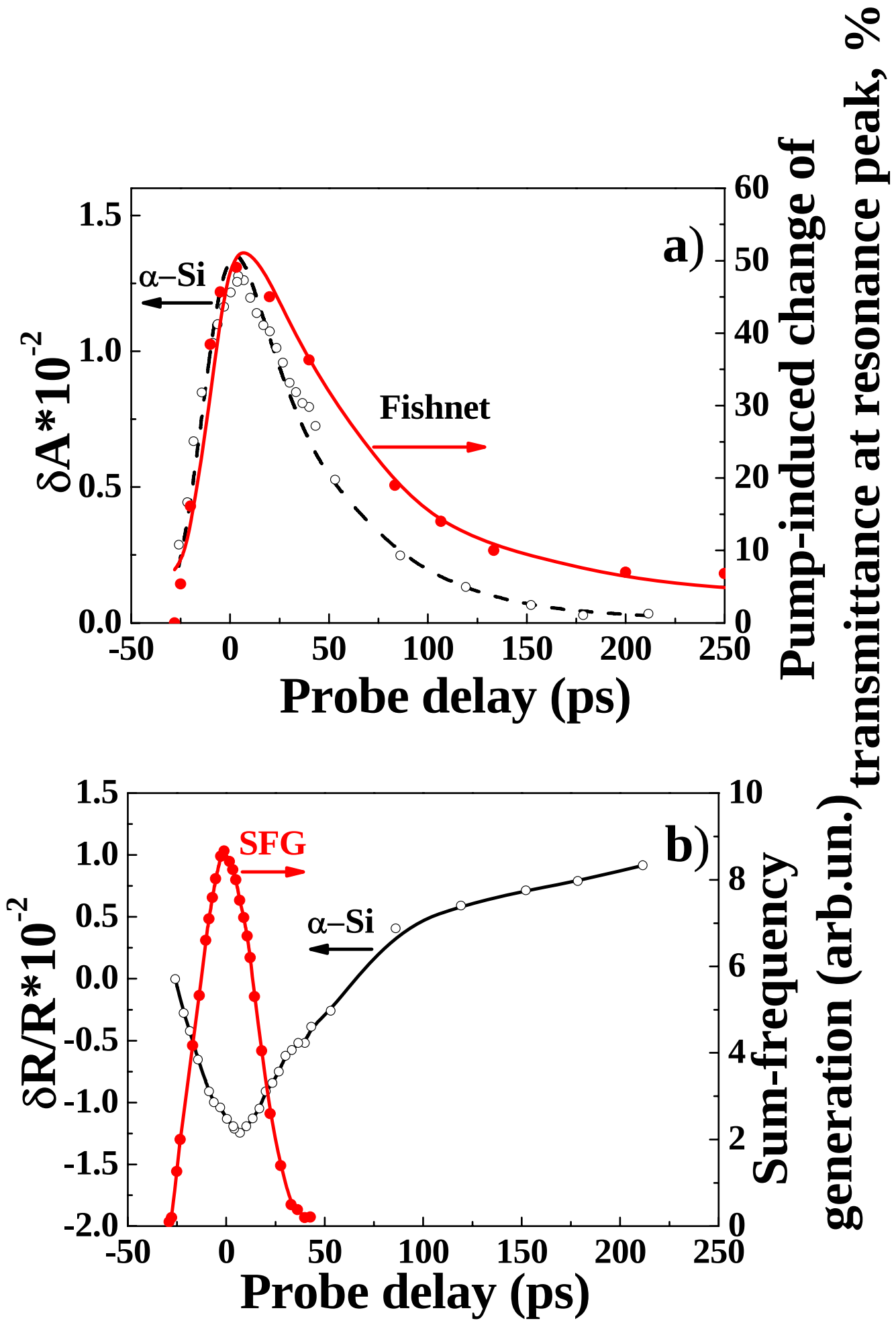

a)
α–Si
Fishnet
δA*10^-2
Pump-induced change of transmitance at resonance peak, %
Probe delay (ps)
b)
SFG
α–Si
δR/R*10^-2
Sum-frequency generation (arb.un.)
Probe delay (ps)

**Figure 3**

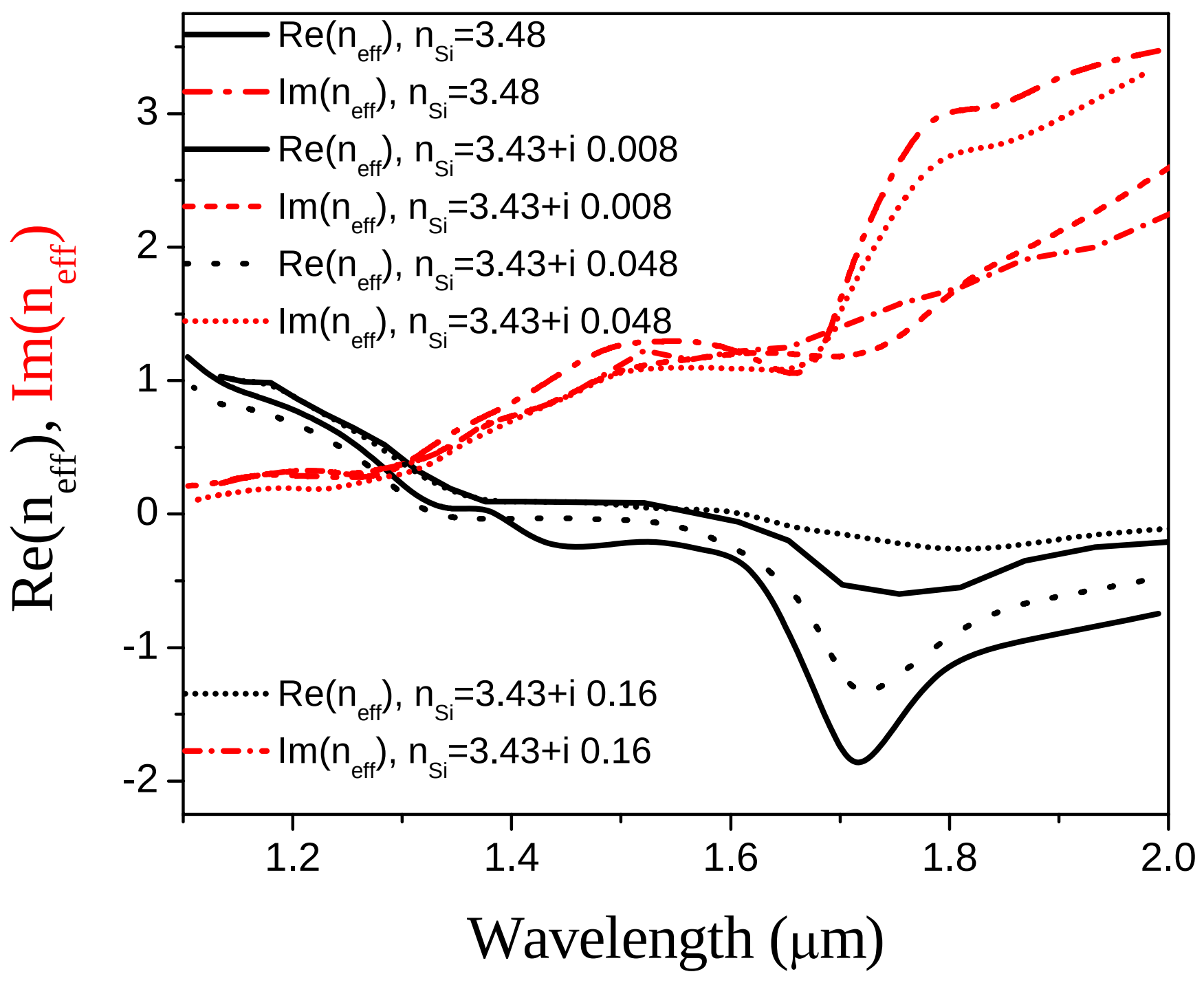

**Figure 4**